\documentclass[reprint,prb,amsmath,amssymb,floatfix,groupedaddress,notitlepage]{revtex4-1}
\usepackage{graphicx}
\usepackage{hyperref}
\usepackage{pdfsync} 
\pdfoutput=1
\makeatletter
\begin{document}
\title{Comment on ``Magnetotransport signatures of a single nodal electron pocket constructed from Fermi arcs"}

\author{Sudip Chakravarty}
\affiliation{Department of Physics and Astronomy, University of
California Los Angeles, Los Angeles, California 90095-1547}
\author{Zhiqiang Wang}
\affiliation{Department of Physics and Astronomy, University of
California Los Angeles, Los Angeles, California 90095-1547}

\date{\today}
\pacs{}
\begin{abstract}
We comment on the recent work [N. Harrison, \textit{et al.} Phys. Rev. B {\bf 92}, 224505 (2015)] which attempts to explain the sign reversal and quantum oscillations of the Hall coefficient observed in cuprates from a single nodal diamond-shaped electron pocket with concave arc segments. Given the importance of this work, it calls for a closer scrutiny.  Their conclusion of sign reversal of the Hall coefficient  depends on a non-generic rounding of the sharp vertices. Moreover, their demonstration of quantum oscillation in the Hall coefficient from a single pocket is  unconvincing. We maintain that at least two pockets with different scattering rates is necessary to explain the observed quantum oscillations of the Hall coefficient.
\end{abstract}
\maketitle

\section{Introduction}
\begin{figure}
\centering
\includegraphics[width=\linewidth]{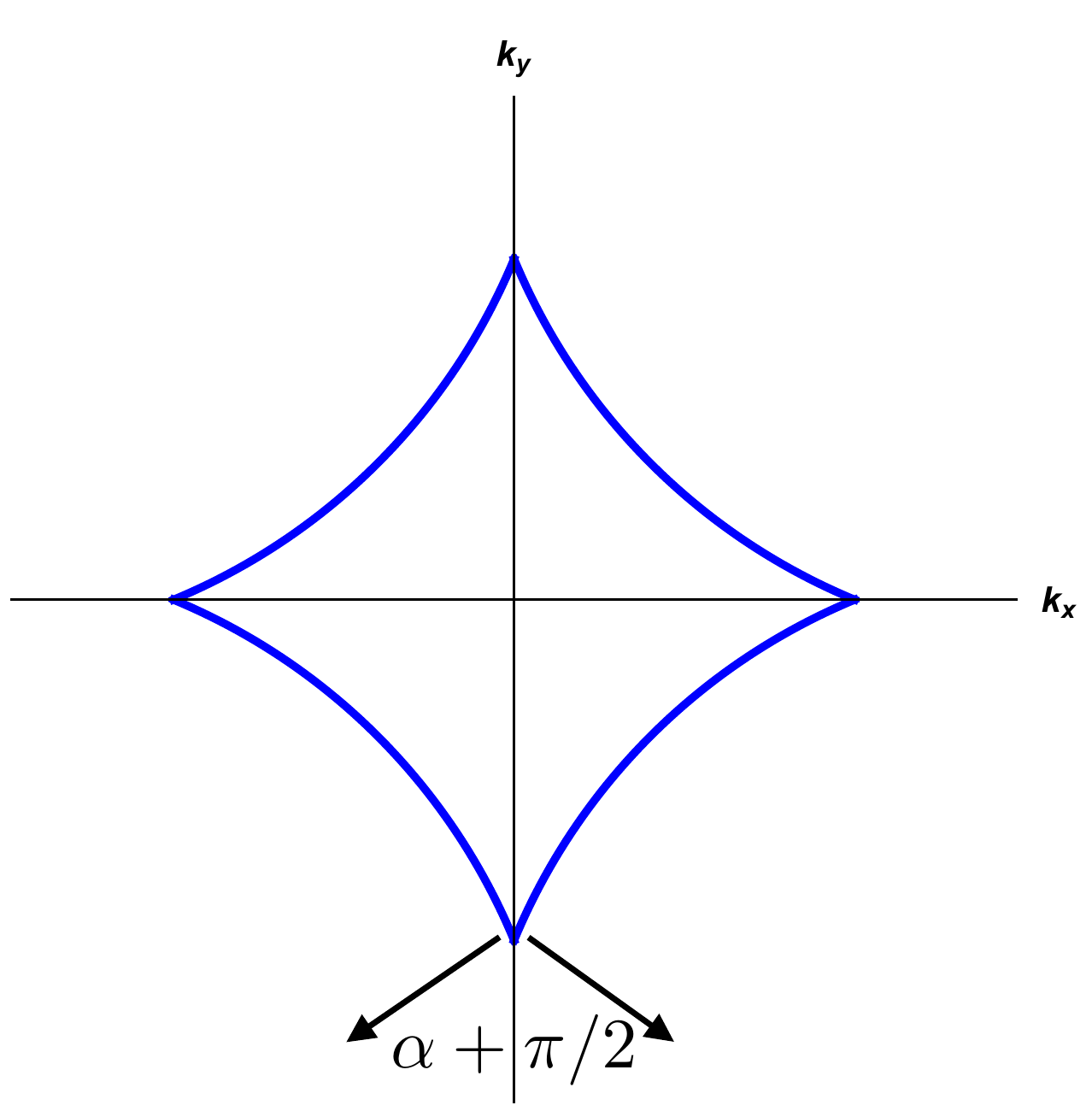}
\caption{Schematic diamond-shaped electron pocket Fermi surface, Ref.~\onlinecite{Harrison2015}.  This pocket is supposed to be centered around the nodal point $(k_x,k_y)=(\frac{\pi}{2 },\frac{\pi}{2 })$ in the Brillouion zone. The lattice spacing is set to unity. When viewed as a part of a circle, each arc has a subtending angle $\alpha$ to the origin of that circle. }
\label{fig: FS}
\end{figure}

In a recent work, \cite{Harrison2015} N. Harrison and S. E. Sebastian  have drawn  attention to  the high-$T_{c}$ community of an intriguing idea.   They have suggested that the sign reversal of the Hall coefficient and its quantum oscillations by making a reconstructed version of the observed Fermi arcs into a single diamond shaped electron pocket shown in Fig~\ref{fig: FS}.   The problem is that any natural reconstruction leads to both electron and hole-like pockets. 

 The nodal Fermi arcs observed in an underdoped cuprate are pieced together  to a single diamond-shaped electron pocket centered at the nodal point $(k_x,k_y)=(\frac{\pi}{2},\frac{\pi}{2})$ of the Brillouion zone, although we do not have an understanding of the Fermi arcs themselves. Given this lack of understanding, the construction by a simple shift of the wave vector needs to be understood in some depth. Be that as it may, we address the simpler aspects assuming that this process can be justified in the future. This electron pocket has four concave arcs  with sharp vertices, as schematically reproduced in Fig~\ref{fig: FS}. Assuming that the magnitude of the velocity is a constant over the whole Fermi surface contour and ignoring any rounding of the sharp vertices, they have been able to compute the magnetic field -$B$- dependent conductivities $\sigma_{xx}$ and $\sigma_{xy}$ by using the semi-classical Shockley-Chambers formula~\cite{Shockley1950,Chambers1952}. Then, from $\sigma_{xx}$ and $\sigma_{xy}$, the Hall coefficient $R_{H}$ can be derived. This   treatment  follows exactly Ref.~\onlinecite{Banik1978}. In Ref.~\onlinecite{Harrison2015}  Harrison and Sebastian have found that if each side of the diamond pocket is sufficiently concave, which means the angle $\alpha$ in Fig~\ref{fig: FS} is large enough, the Hall coefficient $R_{H}$ changes its sign from being positive at $B=0$ to negative at high fields, which can potentially explain the sign reversal of Hall coefficient as a function of  temperature observed in experiments~\cite{LeBoeuf2007}.

To explain the quantum oscillations observed in $R_H$~\cite{Doiron-Leyraud2007}, the authors in Ref.~\onlinecite{Harrison2015}  made the following substitution for the mean free time $\tau$ in the Chambers formula:
\begin{gather}
\tau^{-1} \rightarrow {\tilde{\tau}}^{-1} \equiv [1+2\, \cos(2\pi F/B-\pi) e^{-\pi/\omega_c\tau}]\, \tau^{-1},  \label{eq:subst}
\end{gather}
$F$ is quantum oscillation frequency, and $\omega_c=eB/m^{*}c$, with $m^*$ the cyclotron effective mass, is the cyclotron frequency. By this substitution they obtained quantum oscillations in the Hall coefficient with a single diamond-shaped electron pocket.  It appears to have no justification for oscillations of $R_{H}$.

In this Comment we raise some questions with regard to Ref.~\onlinecite{Harrison2015}. 
On the issue of the sign of the Hall effect  the semiclassical approximation, even though the vertices are a small portion of the Fermi surface,  make a large contribution to $n_H$.  One can describe a limiting process where one starts with a sufficiently smooth rounding of the Fermi surface. Manifestly $n_H$, especially in the large field limit, is simply the enclosed area, and not terribly sensitive to  the shape.  As  the vertices get more singular, the answer  is non-generic.  On the issue of the lack of quantum oscillations for a single pocket, it is generally expected that in the high field limit the Hall  number is exactly given by the enclosed area of the Fermi surface, with no quantum oscillations. This is unfortunate because it is precisely the limit that is relevant. There are also no quantum oscillations in the small field limit because the Dingle factor is exponentially sensitive to $1/B$.  To summarize, unless there is some exchange of particles between multiple bands, or some balance of mobilities between different bands, it is hard to see how  quantum oscillations of $R_{H}$ could arise.

In the following we first show that in the small magnetic field limit, the Shockley-Chambers formula is consistent with the Jones-Zener formula~\cite{Jones1934}, regardless of how the sharp vertices in the Fermi surface contour are rounded. The sign of the zero field Hall coefficient obtained in Ref.~\onlinecite{Harrison2015} heavily relies on a special  rounding the vertices. Different ways of rounding can lead to completely different conclusion about the sign, as correctly pointed out by N. P. Ong in Ref.~\onlinecite{Ong1991}. Therefore the conclusion of the sign reversal of $R_{H}$ obtained in Ref.~\onlinecite{Harrison2015} is not convincing. Finally we explain why the simple replacement in Equation~\eqref{eq:subst} to obtain quantum oscillations is physically  inconsistent. We also give our reasonings why a pronounced quantum oscillation in a Hall coefficient is not expected for a single Fermi surface pocket. At the end we close our comment with some further discussions and conclusion.

\section{Sign of $R_H$ in the weak field limit}
\subsection{ Shockley-Chambers formula  and the Jones-Zener method in the weak field limit}
The Shockley-Chambers formula for the  $2D$  conductivity tensor $\sigma_{\alpha\beta}$ in a magnetic field is~\cite{Ziman1972}
\begin{gather}
\sigma_{\alpha\beta}=\frac{1}{2\pi^2} \frac{e^2}{\hbar} \frac{m^* \omega_c}{\hbar} \int_0^{T_p} d t  \int_0^{\infty} d t^\prime  v_{\alpha}(t) \, v_{\beta}(t+t^\prime)  e^{-t^\prime/\tau} , \label{eq:SC}
\end{gather}
where $\alpha,\beta=x,y$. In this formula the time variable $ t $ (or $t^\prime$) is introduced to parameterize an electron's semi-classical periodic cyclotron motion along the closed Fermi contour under the Lorentz force.  $T_p\equiv 2\pi/\omega_c$ is the time period of one complete circuit motion. The Shockley-Chambers formula is a formal solution to the Boltzman equation in the presence of a perpendicular magnetic field and a longitudinal electric field. This formula itself is applicable in all field regimes, as far as the Landau level quantization effects can be neglected. When these effects are incorporated, there will be quantum corrections to the Shockley-Chambers formula, giving rise to quantum oscillations such as Shubnikov-de Haas effect.

Consider the weak field limit $\omega_c \tau  \ll 1$. Then in Equation~\eqref{eq:SC} we can expand
\begin{gather}
v_{\beta}(t+t^\prime) \approx v_{\beta}(t)+t^\prime \frac{\partial v_{\beta}}{\partial t} \label{eq:expansion}
\end{gather}
because the factor  $e^{-t^\prime/\tau}$  falls off very fast. On the right hand side the first term contributes a zero to the Hall conductivity $\sigma_{xy}$ and therefore to the Hall coefficient $R_H$. The second term gives a contribution $\propto B$ to $\sigma_{xy}$ and therefore a magnetic field independent term to $R_H$ as $R_H\propto \sigma_{xy}/B$ . Higher order terms will give contributions $\sim\mathcal{O}( \omega_c \tau) $ or smaller to the Hall coefficient and therefore vanish in the limit $\omega_c \tau \rightarrow 0$.  In other words in the zero magnetic field limit, the expansion in Equation~\eqref{eq:expansion} becomes exact. For $\sigma_{xx}$, keeping the first term in the expansion of Equation~\eqref{eq:expansion} is enough.

Substituting Equation~\eqref{eq:expansion} into Equation~\eqref{eq:SC} leads to
\begin{align}
\sigma_{xx} & = \frac{1}{2\pi^2} \frac{e^2}{\hbar} \frac{m^* \omega_c \tau }{\hbar} \int_0^{T_p}  v_x(t) v_x(t) d t \, , \\
\sigma_{xy} & = \frac{1}{2\pi^2} \frac{e^2}{\hbar} \frac{m^* \omega_c \tau^2 }{\hbar} \oint v_x(t)  \,  d \, v_y (t)\, ,  \label{eq:SC2}
\end{align}
where the integration path $\oint$ is along the closed Fermi surface contour. Using $\omega_c=eB/m^* c$ and the definition of the magnetic field length $l_B=\sqrt{\hbar c/eB}$ we can rewrite the Hall conductivity as
\begin{gather}
\sigma_{xy} = \frac{e^2}{h} \oint [v_x \tau ] \, d [v_y  \tau]/  \pi l_B^2,
\end{gather}
which is identical to the Jones-Zener method result, see Equation (4) of Ref.~\onlinecite{Ong1991},
\begin{gather}
\sigma_{xy} = \frac{e^2}{h} \oint l_x \, d l_y /  \pi l_B^2,
\end{gather}
if we define a scattering path length vector: $\vec{l}=(l_x,l_y)=(\, v_x(t) \tau(t),v_y(t) \tau(t) \, )$, as in Ref.~\onlinecite{Ong1991}. The assumption here is that $\tau(t) \equiv \tau$ is uniform along the Fermi surface contour.

Therefore within a uniform $\tau$ assumption, the small-field limit of Shockley-Chambers formula  agrees perfectly with the Jones-Zener formula. Note that this conclusion dos not depend on how the sharp vertices in the Fermi surface contour are rounded, contradicting the claim made in Ref.~\onlinecite{Harrison2015} that their consistency do depend on an appropriate rounding of the vertices.

\subsection{Dependence of the sign of $\sigma_{xy}$ on the variation of the Fermi velocity in the vicinity of the vertices }
Although the consistency between the weak field limit Shockley-Chambers formula and the Jones-Zener formula does not depend on how the Fermi velocity around the sharp vertices are modeled, the sign of the computed $\sigma_{xy}(B\rightarrow 0)$ does depend on it crucially~\cite{Ong1991}. Therefore the sign of the $R_H(B\rightarrow 0)$  also heavily relies on the  modeling of the Fermi velocity around the vertex. Different modeling can lead to opposite conclusions about the sign of $R_H(B\rightarrow 0)$.

\subsubsection{The analysis of Banik and Overhauser}
In Ref.~\onlinecite{Banik1978}, Banik and Overhauser defined the Fermi surface piece-wise manner by the four arc segments as in Fig.~\ref{fig: FS}, while neglecting any rounding effects at the  vertices. Assuming that the magnitude of the Fermi velocity  $|\vec{v}_F|=v_F$ is a constant along the Fermi surface contour, the Fermi velocity can be parameterised by
\begin{align}
v_x(t)& =v_F\cos \phi(t) \\
v_y(t) & =v_F\sin \phi(t),
\end{align}
where
\begin{align}
\phi(t)\equiv & \frac{4\alpha}{2\pi} \, \omega_c t - (\frac{\pi}{2}+ \alpha) \sum_{n=1}^{n=4} \theta(\omega_c t - n \pi/2) \nonumber \\
 & +(\frac{\pi}{4}-\frac{\alpha}{2})
\end{align}
is the angle made by the Fermi velocity $\vec{v}_F(t)\equiv (v_x(t),v_y(t))$ with the $x-$axis at time $t$. On the right hand side the second term is a sum of four  step functions, $\theta(x)$. These jumps of $\phi(t)$ at $\omega_c t= n\pi/2$ come from the ``Bragg reflecton" of the particle at each vertex. The initial condition  $ \phi(t=0)=\frac{\pi}{4}-\frac{\alpha}{2}$ has been chosen such that the expression of $\phi(t)$ is simple. According to Equation~\eqref{eq:SC2}, to calculate $\sigma_{xy}(B\rightarrow 0)$ we only need to compute $\tau^2 \oint v_x(t) d v_y(t) $. Because of the discontinuous jumps of $\vec{v}_F(t)$ at $\omega_c t= n\pi/2$ from one side of a vertex to the other, there is a nonzero contribution to the integral $\tau^2 \oint v_x(t) d v_y(t) $ from each vertex. In other words the integral can be decomposed into two parts as follows
 \begin{align}
 &\quad \tau^2 \oint v_x(t) \, d v_y(t) \,\\
& = \tau^2 \left\{\int_{a_1}+\int_{a_2}+\int_{a_3}+\int_{a_4}\right\} v_x  d v_y  \nonumber \\
& \quad -\tau^2 \sum_{n=1}^4 \delta_{\omega_c t, \frac{ n\pi}{2}} \, v_x(t) \lim_{\delta \rightarrow 0} [v_y(t+\delta)-v_y(t-\delta)]\\
& \equiv A_{a} - A_{d}.
\end{align}
On the second line, $\int_{a_1}, \int_{a_2}, \int_{a_3},\int_{a_4}$ stand for integrations along the four arc segments in Fig.~\ref{fig: FS}. The sum of these four terms is denoted as $A_a$ in the last line. The subscript ``$a$" in $A_a$ stands for ``arc".  The third line is a sum of the discontinuous contributions of the Fermi velocity  from the four vertices, as indicated by the Kronecker-delta $\delta_{\omega_c t, \frac{ n\pi}{2}}$.  This sum is then denoted as $A_d$ in the last line, where the subscript $d$ stands for ``discontinuity", stressing that it comes from the discontinuous jumps of the Fermi velocity.

A little inspection shows that $A_a$ is equal to the sum of the Stokes area swept out by the scattering path vector $\vec{l}$ as an electron moves along each Fermi arc segment. Therefore
\begin{gather}
A_a=4 \frac{1}{2} \alpha (v_F\tau)^2=2 \alpha l^2,  \label{eq:Aa}
\end{gather}
where $l=v_F\tau$ is the mean free path.

Computation of $A_d$ is straightforward and the final result is
\begin{gather}
A_d= 2 l^2 \cos \alpha.
\end{gather}
Similar to $A_{a}$, $A_d$ also has a geometric interpretation. This can be seen clearly if we anti-symmetrize $v_x(t)$ and $v_y(t)$ in calculating the Hall conductivity $\sigma_{xy}$. After the anti-symmetrization $A_d$ can be rewritten as follows
\begin{gather}
\frac{A_d}{\tau^2}= \sum_{n=1}^4 \delta_{\omega_c t, \frac{ n\pi}{2}} \,  \frac{1}{2} \left[ v_x(t) \Delta v_y(t)-v_y(t) \Delta v_x(t)\right],
\end{gather}
where $\Delta v_{x/y}(t)=\lim_{\delta \rightarrow 0} [v_{x/y}(t+\delta)-v_{x/y}(t-\delta)]$. Now each term in the above sum can be identified as the area of the triangle made by the two Fermi velocity vectors on the two sides of each vertex. Each of them is equal to $\frac{1}{2} v_F^2 \sin(\frac{\pi}{2}+\alpha)=\frac{1}{2} v_F^2 \cos \alpha $. Therefore the sum is equal to $2 v_F^2 \cos \alpha$ and $A_d=2 l^2 \cos \alpha$. Hence the Hall conductivity $\sigma_{xy} \propto (A_a-A_d) \propto (\alpha-\cos \alpha)$. Correspondingly the Hall coefficient $R_H \propto (\alpha -\cos \alpha)$. So it changes sign as $\alpha$ changes from $\pi/2$ to $0$.

From this analysis we see that zero field Hall conductivity  contains not only a contribution $A_a$ from each arc segment, but also another contribution $A_d$ from the discontinuous jumps of the Fermi velocity from one arc to the adjacent arc at each vertex. We should emphasize that this $A_d$ contribution exists without taking into account how the vertices are rounded.

\subsubsection{Harrison  and Sebastian rounding of the vertices}
After computing the $\sigma_{xy}$ directly from the Shockley-Chambers formula, the authors of Ref.~\onlinecite{Harrison2015} then tried to calculate the $\sigma_{xy}$ by computing the Stokes area swept out by the scattering path vector as an electron moves along the entire Fermi surface contour, following N. P. Ong~\cite{Ong1991}. There are two contributions: one from the four disjoint arc segments; the other from the vicinity of vertices. The arc segment contribution is equal to $A_a$ as computed in the previous section and given by Equation~\eqref{eq:Aa}. Computing the vertex contributions requires  a knowledge of how the sharp vertices are rounded and how the Fermi velocity varies near the vertices after rounding. We denote this contribution as $A_v$, where the subscript ``$v$" stands for vertices. Then the authors of Ref.~\onlinecite{Harrison2015} have chosen a special way of modeling the vertices and computed $A_v$. The surprising thing is that the $A_v$ they have computed is  identical to $A_d$ introduced in the previous section, which comes from the discontinuous jump of the Fermi velocity at the vertices without rounding. Therefore the $\sigma_{xy}\propto A_a - A_v$ calculated in Ref.~\onlinecite{Harrison2015} is identical to the $\sigma_{xy} \propto A_a-A_d$ computed in the previous section by following Banik and Overhauser. Based on this fact the authors of Ref.~\onlinecite{Harrison2015} have claimed that such an agreement shows that they have appropriately modeled the variation of the Fermi velocity in the vicinity of vertices  using Ref.~\onlinecite{Banik1978} to compute the $\sigma_{xy}$. But from our analysis we see that this claim cannot be true. The agreement between $A_v$ and $A_d$ they have found is  a coincidence, not generic. A different way of rounding the vertices can give a contribution $A_v$ that is completely different from $A_d$ in general;  see Ref.~\onlinecite{Ong1991}.  
In short, the sign of the zero field $\sigma_{xy}$  depends on how the sharp vertices are rounded. The special modeling of the vertices in Ref.~\onlinecite{Harrison2015} might be  artificial. Therefore it puts the conclusion obtained about the sign of the low field $\sigma_{xy}$ in  doubt. A slightly more realistic modeling of the Fermi velocity around the vertices might change the final conclusion.

\section{High magnetic field regime}
In the following we first give our reasons why a simple replacement of $\tau$ with $\tilde{\tau}$ in the Shockley-Chambers formula to extract quantum oscillation is wrong and also give our arguments why we do not expect pronounced quantum oscillations of the Hall coefficient from a single Fermi surface pocket. 

\subsection{Inconsistency of the replacement of $\tau^{-1}$ with $\tilde{\tau}^{-1}$ in Schokley-Chambers formula}
To obtain quantum oscillations in the Hall coefficient the authors in Ref.~\onlinecite{Harrison2015} made a simple substitution of the mean free time $\tau$ in Equ.~\eqref{eq:subst} into the Schokley-Chambers formula in Equ.~\eqref{eq:SC}. However this kind of treatment can not be correct. We know that the Schokley-Chambers formula is a formal solution to the semi-classical Boltzman equation in the presence of both an electric field $\vec{E}$ and a perpendicular magnetic field $\vec{B}$
\begin{gather}
(-e) \vec{E}\cdot \vec{v}_{\vec{k}} \, \frac{\partial f^0}{\partial \epsilon}+ \frac{-e}{\hbar c} \vec{v}_{\vec{k}} \times \vec{B} \cdot \nabla_{\vec{k}} \, g =\frac{g}{\tau},
\end{gather}
where $f^0$ is the equilibrium distribution in the absence of fields $\vec{E},\vec{B}$ and $g$ is the out of equilibrium distribution due to the fields. The total distribution is $f=f^0+g$. The right hand side of this equation accounts for the relaxation back to the equilibrium distribution due to incoherent scattering processes within the relaxation time approximation.
The  replacement of $\tau^{-1}$ in the Boltzman equation with $\tilde{\tau}^{-1}$ is hard to justify. 
\subsection{Arguments disfavoring pronounced quantum oscillations of Hall coefficient from a single Fermi surface pocket}
We believe that a single Fermi surface pocket is unlikely to give pronounced quantum oscillations in the Hall coefficient. Our claim is based on two extreme considerations. First consider the weak field limit $\omega_c \tau \rightarrow 0$. In this limit because the field is too weak $\omega_c \ll 1/\tau$ any Landau level quantization effects is washed out by disorder scattering effects. So no quantum oscillation can be observed. Next consider the high field limit $\omega_c \tau \gg 1$.  In the presence of a longitudinal electric field and a perpendicular magnetic field, we know semi-classically the motion of an electron will be a cyclotron motion superimposed on top of a uniform drift. In the $\omega_c \tau \rightarrow \infty$ limit, the drift motion completely dominates over the cyclotron orbit motion so that ~\cite{Ashcroft1976}
\begin{gather}
\lim_{\omega_c \tau \rightarrow \infty} \vec{j}_{\perp}=- n e \vec{w}=-\frac{n e c}{B} \vec{E}\times\hat{B}.
\end{gather}
Here $\vec{j}_{\perp}$ is the current density in the direction perpendicular to both the electric field and the magnetic field. $n$ is the density of charge carriers. $\vec{w}=c\frac{E}{B}\hat{E}\times \hat{B}$ is the drift motion velocity. Therefore in this limit the Hall coefficient is simply $R_H=|\vec{E}|/(|\vec{j}_{\perp}| B)=-\frac{1}{n e c}$ for electron like carriers and becomes field independent. In the quantum mechanical picture, the electron's motion can still be decomposed into a drift motion of the center and a quantized cyclotron orbital motion around the center as long as the semiclassical orbits are closed. Therefore  the conclusion remains the same. Hence in the high field limit no quantum oscillation exist in Hall coefficient. Then by interpolation we do not expect any pronounced quantum oscillations of the Hall coefficient from a single Fermi surface pocket observed at some intermediate value of $\omega_c \tau$.


\section{Conclusion}
The conclusion about the sign of the zero field Hall conductivity/coefficient obtained in Ref.~\onlinecite{Harrison2015} heavily relies on a special modeling of the sharp vertices in the diamond-shaped electron pocket Fermi surface of Fig.~\ref{fig: FS} and is therefore non-generic. The quantum oscillation in the Hall coefficient obtained in Ref.~\onlinecite{Harrison2015} was based on an inconsistent substitution of the mean free time with an oscillatory mean free time in the Shockley-Chambers conductivity formula. We have give our own arguments disfavoring a pronounced quantum oscillation in Hall coefficient from a single Fermi surface pocket. 

 The negative Hall coefficient is quite a general result in the cuprates. Even in cuprates where it's not negative at higher temperatures, it  heads towards negative values at low temperatures. The specifics of the CDW, on the other hand, vary quite a bit between different cuprates. And one could imagine that the details of the rounding of the corners  would be very different indeed. Therefore it seems unlikely that something so general--negative Hall coefficient--could rely on something so specific---corner rounding.

YBCO is indeed in the crossover regime of $\omega_c \tau \sim1$ regime. However,  even if the substitution $\tau \to \tilde \tau$ was correct, it would necessarily lead to decreasing quantum oscillation amplitude in the Hall channel with field, as the high field regime is approached, where the hall effect is purely geometrical. This is clearly not observed in experiments.

\begin{acknowledgements}
We thank Ching-Kit Chan, S. Kivelson, B. J. Ramshaw and I. Vishik for many discussions.  This work was partly performed at the Aspen Center for Physics, which is supported by National Science Foundation grant PHY-1066293. It was also supported in part by the funds from the David S. Saxon Presidential Term Chair at UCLA.
\end{acknowledgements}

\end{document}